\documentclass{raa}         
\usepackage{graphicx,times}
\usepackage{natbib}
\usepackage{amssymb,amsmath}
\bibpunct{(}{)}{;}{a}{}{,}

\usepackage[a4paper=true,dvipdfm=true,pagebackref=true]{hyperref}
\hypersetup{pdftitle = The title of my PDF, pdfauthor = My name, pdfsubject= The subject, pdfkeywords = keyword1 keyword2 keyword3} 
\hypersetup{colorlinks = true, linkcolor = green, anchorcolor = red, citecolor = blue, filecolor = red, pagecolor = red, urlcolor = red}

\usepackage{xcolor}

\begin{document}
\title{The intrinsic SFRF and sSFRF of galaxies: comparing SDSS observation with IllustrisTNG simulation}

 \volnopage{ {\bf 20XX} Vol.\ {\bf X} No. {\bf XX}, 000--000}
   \setcounter{page}{1}

   \author{Ping Zhao\inst{1}, Haojie Xu\inst{1}, Antonios Katsianis\inst{2, 1},  Xiaohu Yang
      \inst{1, 2}
   }

   \institute{Department of Astronomy, School of Physics and
  Astronomy, Shanghai Jiao Tong University, Shanghai, 200240, China; {\it haojie.xu@sjtu.edu.cn; kata@sjtu.edu.cn} \\
        \and
             Tsung-Dao Lee Institute, and Shanghai Key Laboratory
  for Particle Physics and Cosmology, Shanghai Jiao Tong University,
  Shanghai, 200240, China \\
\vs \no
   {\small Received 2020-04-05; accepted 2020-05-30}
}

\abstract{The star formation rate function (SFRF) and specific star formation rate function (sSFRF) from the observation are impacted by the Eddington bias, due to the uncertainties on the estimated SFR. We develop a novel method to correct the Eddington bias and obtained the intrinsic SFRF and sSFRF from the Sloan Digital Sky Survey Data Release 7.
The intrinsic SFRF is in good agreement with measurements from previous 
data in the literature that relied on UV SFRs but its high star-forming end is slightly lower than those IR and radio tracers. We demonstrate that the intrinsic sSFRF from SDSS has a bi-modal form with the one peak found at ${\rm  sSFR \sim 10^{-9.7} yr^{-1}}$ representing the star-forming objects while the other peak is found at ${\rm  sSFR \sim 10^{-12} yr^{-1}}$ representing the quenched population. Furthermore, we compare our observations with the predictions from the IllustrisTNG and Illustris simulations and show that the 
``TNG'' model performs much better than its predecessor. However, we show that the simulated SFRF and cosmic star formation density (CSFRD) of TNG simulations are highly dependent on resolution, reflecting the limitations of the model and today state-of-the-art simulations. We demonstrate that the bi-modal, two peaked sSFRF implied by the SDSS observations does not appear in TNG regardless of the adopted box-size or resolution. This tension reflects the need for inclusion of an additional efficient quenching mechanism to the TNG model.
}

   \authorrunning{P. Zhao et al. }            
   \titlerunning{The SFRF and sSFRF in SDSS and in IllustrisTNG}  
   \maketitle

%
\section{Introduction} 
\label{sect:intro}
The Star Formation Rate (SFR) taking place in galaxies and across the Cosmos represents a fundamental constraint for galaxy formation physics and stellar evolution models. The number density of star-forming galaxies as a function of their star formation rate, i.e. the Star Formation Rate Function (SFRF) provides qualitative and quantitative information about star formation occurring in galaxies, while by definition its integration results in the Cosmic Star Formation rate Density (CSFRD).

To obtain the SFRs of galaxies, observational studies typically have to rely on models which provide correlations between SFR and the observed Ultra-Violet (UV) \citep{Santini2017,Blanc2019}, Infra-Red (IR) \citep{Whitaker2014,Guo2015}, H$\alpha$  \citep{Cano2019}, O[II] emission \citep{Lopez2020} and Radio luminosities  \citep{Karim2011} or the SED fitting method \citep{Duncan2014,Kurczynski2017,Trcka2020}. Some studies in the literature rely on more than one indicator/methodology to provide a multi-wavelength analysis \citep{Davies2019,Katsianis2019}, with some finding discrepancies between the different techniques \citep{Davies2016,Katsianis2016} and others not \citep{Madau2014,Driver2016}. Nevertheless, most of the studies in the literature acknowledge that every single methodology has advantages but at the same time shortcomings \citep{2009ApJ...706..599L,Katsianis2020}. For example, UV light is subject to dust attenuation effects,\citep{Dunlop2017,Baes2020} and usually not complete for bright high star-forming galaxies. It provides information for intermediate and low star-forming galaxies at high redshifts (z $>$ 2) but is not that successful at the lower redshifts \citep{Katsianis2017}. On the other hand, the IR luminosity originating from dust continuum emission is a good tester of dust physics \citep{2003A&A...410...83H, Katsianis2015} with IR wavelengths (especially Mid-IR and Far-IR) being used to determine the total IR luminosity. Severe drawbacks of IR studies though is that a) they usually do not have sufficient wavelength coverage \citep{Lee2013,Pearson2018}, b) can be compromised by Active Galactic Nuclei \citep[AGN][]{Roebuck2016,Brown2019}, c) have to rely on SED libraries \citep{DaleHelou2002,Wuyts2008}, which have been constructed  from  galaxies  at  low redshifts and are not reliable at higher redshifts, d) other sources can contribute to the heating of dust in galaxies and this contribution can be falsely taken as star formation, for example, old stellar populations can significantly contribute to dust heating, complicating the relation between SFR and TIR emission \citep{Viaene2017,Nersesian2019}. Except for UV and IR light, H$\alpha$ photons can be used to trace the intrinsic SFRs too. However, H$\alpha$ light is subject to severe dust attenuation effects, can usually probe intermediate star-forming objects \citep{Katsianis2016}, and usually incomplete for the high star-forming systems. Due to the above limitations of SFRs derived from monochromatic luminosities other studies employ the SED fitting techniques to numerous bands \citep{Leja2019,Hunt2019}. However, \citet{Katsianis2014} and \citet{Santini2017} suggested that this method suffers from parameter degenerations, which are serious for the SFR estimation. Besides the fact that star formation rate represents an excellent and direct instantaneous census of star formation, most articles instead of focusing on the star formation rate function they usually study the stellar mass function which involves an integrated property with time.

Cosmological simulations are a valuable tool to study galaxy formation since the story of the Universe is of high complexity involving different astrophysical processes, like the non-linear evolution of dark matter halos, feedback, gas heating/cooling, and chemical processes. Cosmological-scale simulations such as Illustris \citep{Vogelsberger2014}, Blue  Tides  \citep{Feng16},  Horizon-AGN \citep{Kaviraj20117}, Mufasa \citep{Dave2017}, Romulus  \citep{Tremmel2017}, IllustrisTNG  \citep{Springel2018,Pillepich2018} and SIMBA \citep{Dave199} use sub-grid  models to  reproduce stellar, gaseous, and black hole components that attempt to resemble those in observed  galaxies. Besides, Semi-analytic models like the Durham model \citep{Cole2000}, L-GALAXIES \citep{Guo2013}, GALACTICUS \citep{Galacticus14} and SHARK \citep{Shark2019} have allowed studying galaxy formation in larger volumes. More specifically, the evolution of the star formation rate function has been studied by some hydrodynamic simulations \citep{Dave2011,TescariKaW2013,Katsianis2016,Dave2017,Canas2019} and semi-analytic models \citep{Fontanot2012,Gruppioni2016} at different redshifts. \citet{TescariKaW2013} and \citet{Katsianis2016} demonstrated the importance of feedback from SNe and AGN in the evolution of the star formation rate function for z $\sim1-7$ galaxies. \citet{Gruppioni2016} compared semi-analytic models \citep[e.g.][]{Monaco2007,Henriques2015} with IR observations. The comparison showed that semi-analytic models under-predict the bright end of the SFRF at intermediate and high redshifts. \citet{Dave2017} compared Mufasa to observed  galaxy SFRs and sSFRs. At $z=0$, the simulated SFRF is in good agreement with \cite{bothwell2011} but has higher normalization by up to $\sim\times 3$ in comparison with the \cite{Gunawardhana2013} data from the Galaxy And Mass Assembly (GAMA) survey. The authors also compared the simulated specific star formation rate functions (sSFRs) with the observed sSFRF given by \citet{Ilbert2015} demonstrating a good agreement in most stellar mass bins. Last, \citet{Katsianis2017} demonstrated that the  SFRF of the EAGLE reference simulation is in good agreement with the UV and H$\alpha$ observations at $z=0$, while distributions that originate from IR and Radio data suggest a higher number density of high star-forming systems. The authors demonstrated that the reason for this inconsistency is the presence of the AGN feedback in EAGLE, which is though important at reproducing the UV and H$\alpha$ data. 

 The Sloan Digital Sky Survey Data Release 7 (SDSS DR7; \citealt{York00, Strauss02, Stoughton02, Abazajian09}) is one of the most successful and well-studied galaxy redshift surveys for the local Universe. Its spectroscopic nature enables the accurate redshift and well inferred stellar mass and SFR for more than half of millions of galaxies. Therefore, SDSS DR7 provides a good opportunity to construct star formation rate functions, specific star formation rate functions, and cosmic star formation rate densities for the local Universe \citep{Yang2013}. The above can be compared with previous studies that employed different SFR indicators and techniques and provide further constraints to cosmological simulations and semi-analytic models. Besides, Illustris and its successor IllustrisTNG represent two state-of-the-art cosmological hydrodynamic models that have been successful at reproducing numerous observations. It would be interesting to perform a direct comparison with the observed SFRFs and sSFRFs from the observations, and point out any agreements or inconsistencies.
 
 The observed SFRF, due to the uncertainties on star formation rate estimation, 
inevitably suffers from the so-called Eddington bias \citep{Eddington1913}. 
The Eddington bias simply describes the fact that when counting the 
number of the galaxies in bins of galaxy properties (e.g., luminosity, stellar mass, SFR, and host halo mass), errors in the estimation of the properties leads to potential biases to the histograms (e.g., luminosity function; stellar mass function \citep{Caputi2011, Ilbert2013} or halo mass function \citep{Dong2019}). The extent of the Eddington bias depends on the size of errors and the shape of the histograms. For instance, at the exponential cutoff part, there will be significantly more galaxies scattering from lower bins to higher ones than the reverse, which severely biases the density of luminous/massive galaxies. In the context of SFRF, it would be expected that the density of high star-forming galaxies is overestimated. Therefore, using the observed Eddington-biased SFRF directly computed from the observations would prevent us from a fair comparison with the predictions from cosmological simulations, especially at the high star-forming end.
 
The structure of the paper is as follows: In section~\ref{sect:Obs}, we will present and test our methodology for correcting the Eddington bias on SFR function
and infer the intrinsic star formation rate functions
and specific star formation rate functions for the SDSS DR7. In section~\ref{sect:Comp}, we introduce briefly the Illustris and IllustrisTNG suite of simulations and compare these with the SFRFs and sSFRFs from SDSS DR7. We summarize and discuss our conclusions in section~\ref{sect:Res}. Throughout the work, we adopt a spatially flat $\Lambda$ cold dark matter cosmology with $\Omega_{\rm m}=0.275$ (WMAP7; \citealt{Komatsu2011}) to convert the redshift to comoving distance. To facilitate fair comparisons on the SFRFs and sSFRFs, we convert using the corresponding Hubble constants adopted by the various simulations and observations employed in this work. We use $\log$ for base-10 logarithm.



\section{A method for correcting the Eddington Bias on the Star Formation Rate Function and Specific Star Formation Rate Function}
\label{sect:Obs}

The galaxy properties (e.g., magnitude and redshift) used in this work are
obtained from the New York University Value-Added Galaxy catalog (NYU-VAGC; \citealt{Blanton2005}). 
We adopt the star formation rates
specific star formation rates, and their uncertainties provided by the MPA-JHU group\footnote{\url{https://wwwmpa.mpa-garching.mpg.de/SDSS/DR7/sfrs.html}}.
The star formation rates are computed by fitting the emission lines 
(e.g., H$\alpha$, H$\beta$, [O III]5007, [N II]6584, [O II]3727, and [S II]6716) with
Bayesian methodology and model grids (see details in \citealt{Brichram04}).
The stellar masses of galaxies are taken from \citet{Kauffmann2003}, who estimated these by
using two stellar absorption-line indices, 
the 4000\AA$~$break strength, and the Balmer absorption-line index H$\delta_{A}$.
The specific star formation rates are simply calculated by combining the star formation rate and
stellar mass likelihoods aforementioned. Through this work, we take the median values 
from the SFR/sSFR posterior probability distributions as our best values. 
With taking account cases with asymmetric probability distributions, we estimate the uncertainties as the mean 
34 percentiles from the median given 16th and 84th percentiles of probability distributions.
In detail, these uncertainties are mostly from degenerations produced in the SED fitting and also include different sources of errors, such as the photometric errors, the wavelength coverage, and the limited SED template grids.
A \citet{kroupa01} initial mass function is assumed in the derivation of the quantities.
All the source data we used in this work are compiled here\footnote{\url{http://gax.sjtu.edu.cn/data/Group.html}}.

\subsection{The Eddington Bias Correction on Star Formation Rate Function and Beyond}
\label{subsec:EB_correction}

\begin{figure} 
   \centering
   \includegraphics[width=0.7\textwidth]{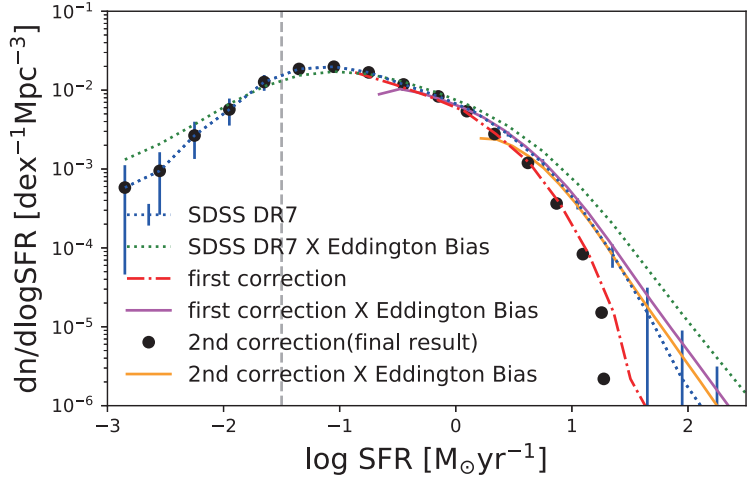}
   \caption{The Eddington bias correction on the star formation rate function in SDSS DR7. The blue (green) dashed line represents observed SFRF computed from SDSS DR7 (observed SFRF convolving with SFR uncertainties). The red dash-dotted line (black dots) is/are the first (second) correction, and the magenta (orange) solid line is the first (second) correction plus Eddington bias.
   Based on the definition of Eddington bias, the second correction is
   ought to be the intrinsic SFRF we are seeking. The flattening
   behavior in the left side of first/second correction plus Eddington bias 
   line is totally artificial since when plotting, we only build the histogram with galaxies with updated SFRs.
   If taking all galaxies into account, both left sides will line on
   top of SDSS DR7 (see magenta and yellow solid lines in Fig. \ref{fig:simulation_proof}).
   See details in section ~\ref{subsec:EB_correction}. The error bars on the observed SFRF 
   are computed from $150$ jackknife samples \citep{Xu2016, Xu2018}.
   The grey vertical dashed line marks the complete boundary for the observations and simulations, and on its right forms the analysis of this work.
   } 
   \label{fig:EBcorrection}
   \end{figure}

In this subsection, we present and test our method for correcting the 
Eddington bias in star formation rate function (SFRF). We start with the 
observed SFRF computed using the so-called 1/Vmax  weighting method \citep{1976ApJ...207..700F, Li2009} to correct the
Malmquist bias due to the flux-limited survey nature.
We note that Vmax is calculated from r-band Petro magnitude 
 (K+E corrected to $z$=0.1), with also spectroscopy completeness 
 taken into account.

Since the observed SFRF is a result of convolution between the intrinsic SFRF and 
the uncertainties on star formation rates, in principle, we can 
get rid of the Eddington bias by seeking a function 
that after convolving with the provided star formation rate uncertainties the resultant function matches the observed SFRF.
Motivated by this idea, we develop an empirical method to 
remove the Eddington bias 
by the following steps and each step is also illustrated in Fig. \ref{fig:EBcorrection}.

(i) Step 1: We convolve the observed SFRF with the SFR uncertainties 
to obtain a function
(dubbed ``SDSS DR7 $\times$ Eddington bias'' in Fig. \ref{fig:EBcorrection} ). 
This new function can be understood as a 
SFRF observed in a world contaminated by the Eddington bias {\it twice}. 
Here and after, by convolving SFRF (or stellar mass function in the later section) 
with SFR (stellar mass) uncertainties, in practice, we draw 1000 SFRs (stellar masses) 
for each galaxy with the assumption that
each galaxy follows a Gaussian distribution around its median SFR (stellar mass) and  
its uncertainty is the standard deviation. We then build a histogram of 1000 mocks and take the median value in each bin as the bin value. 
By doing this, we effectively inject the Eddington bias effect.

(ii) Step 2: After recording down the difference in the x-axis ($\log$ SFR) between 
the observed SFRF and ``SDSS DR7 $\times$ Eddington bias'' as a function of x-coordinates
of observed SFRF, we interpolate and apply the ``correction'' to
the individual galaxy SFRs to effectively remove the Eddington bias on
the level of an individual galaxies. 
With updated SFRs for all the galaxies, we are ready to build the histogram of a new
SFRF labeled ``first correction'' SFRF. 
We note that the galaxies with SFR 
$\leq 10^{-0.2}$ ${\rm M_{\odot} yr^{-1}}$ by design do not need to be corrected for
Eddington bias, while the star-burst galaxies (SFR 
$\geq$ 10 ${\rm M_{\odot}yr^{-1}}$) require a considerable correction.

(iii) Step 3: We then convolve the ``first correction'' SFRF with the SFR 
uncertainties again to obtain the 
so-called ``first correction $\times$ Eddington bias'' SFRF function.
For simplicity, we assume that the galaxy SFR uncertainties
stay the same regardless of their change on SFRs.

(iv) Step 4: If the ``first correction $\times$ Eddington bias'' function 
match with the observed SFRF, it implies 
that the ``first correction'' should be the
intrinsic Eddington-bias-free SFRF. If not, as our SFRF case shown
in Fig.~\ref{fig:EBcorrection}, we go back to Step 2 by recording
the difference in x-axis between the observed SFRF and 
{\it ``first correction $\times$ Eddington bias'' function} 
as a function of x-coordinates
of observed SFRF, we interpolate and apply this additional ``correction'' to
the already updated SFRs of all galaxies. We then plot the ``2nd correction'' SFRF
with the twice updated SFRs. Note that the first (second, ..., Nth) corrections
that applied to the individual galaxy SFRs should always come from
the difference in x-axis between the observed SFRF and 
{\it ``first (second, ... , Nth) correction $\times$ Eddington bias " function} as a function of x-coordinates of observed SFRF. These corrections should be asymptotic to 0, 
as the ``first (second, ... , Nth) correction $\times$ Eddington bias function''
converges to the observed SFRF.

(v) Step 5: We repeat the step 3 but convolve the ``2nd correction'' SFRF with the SFR uncertainty. After that, repeat step 4 and 5 until the ``N-th correction $\times$ Eddington bias'' function matches the observed SFRF. 

   
The intrinsic SFRF can be found by iteratively applying these steps until
the ``N-th correction X Eddington bias'' function matches the observed SFRF. 
For our SFRF case, it only takes us 2 iterations to arrive at the
Eddington-bias-free SFRF, shown in Fig.~\ref{fig:EBcorrection}. We note that by no means we declare our method as an exact method for
recovering the intrinsic SFRF due to many approximations and 
simplifications used in the assumptions and detailed procedures. 
However, we believe that, to the zeroth-order correction, the function inferred 
from this method ought to be much
closer to the intrinsic SFRF than the observed SFRF.

To ensure that our method recovers (or at least approaches as close as possible to) 
the intrinsic SFRF, 
we test our method in the TNG100-1 simulation in Appendix \ref{appendixilleb}. In short, our method for correcting the Eddington bias in SFRF
is proven to work as expected in the simulation, with the simplest 
configuration though. The test gives us strong confidence in our inference
on the SDSS DR7 intrinsic SFRF. In principle,
our method can be applied to any Schechter-like histogram, 
such as luminosity function and stellar mass function.
Compared to the previous work on correcting the 
Eddington bias on stellar mass function \citep{Caputi2011, Ilbert2013}, 
our method has much more flexibility since we do not assume a functional form for the stellar mass
function (we also apply our method to the stellar mass function in Appendix. \ref{subsec:SMF_EB_appendix}). The intrinsic 
SFRF of the local Universe is listed in 
Table~\ref{tablel} and also shown as black points in Fig.\ref{Fig1}.

\subsection{The Intrinsic Star Formation Rate Function}

\begin{figure} 
   \centering
   \includegraphics[width=0.7\textwidth]{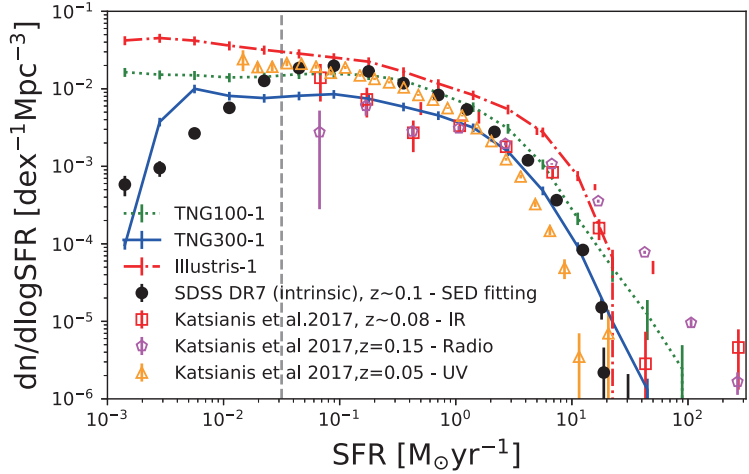}
   \caption{The star formation rate functions of the local Universe: The green dotted line represents the SFRFs derived from TNG100, the blue solid line from TNG300-1, and the red dash-dotted line from Illustris-1. The black filled dots represent the intrinsic SFRFs inferred from SDSS DR7 (Eddington bias removed, see text for details), the empty red square, magenta pentagon and orange triangle shows observed SFRFs obtained via tracers as IR, Radio and UV LFs, respectively. The error bars on SDSS DR7 SFRF are 
   obtained by the 150 Jacknife subsamples.
   } 
   \label{Fig1}
   \end{figure}

In Fig.~\ref{Fig1} we present a comparison between our results from SDSS DR7 with the SFRFs given by \citet{Katsianis2017}.  We demonstrate that the intrinsic SFRF from our analysis is overall in good agreement with the SFRF derived from the UV luminosity function of \citet{Robotham2011}, especially when the comparison is made below the SFR limit of 5${\rm M_{\odot}yr^{-1}}$. However, at the high star-forming end ($ {\rm SFR} > 10  {\rm M_{\odot} \, yr^{-1}}$), the intrinsic SFRF lays between the SFRFs obtained from the UV data and the IR data from \citet{Patel13}. As mentioned in the introduction, UV light is subject to dust attenuation effects. This usually makes UV studies incomplete at the bright end since high star-forming objects with huge contents of dust will not be present in the survey. Besides, since dust attenuation effects become more severe for high star-forming objects, any applied dust corrections to infer the intrinsic SFRs can be underestimated \citep{meurer1999,Katsianis2020}. Both effects can result in underestimated SFRFs at the high star-forming end from UV data. On the other hand, IR light can be enhanced by other sources (e.g. old stellar populations, Active Galactic Nuclei) and this augmentation can be falsely taken as additional star formation, especially in massive/old galaxies. The above can result in overestimated IR SFRFs at the high star-forming end. The SED derived SFRF from SDSS DR7 lays between the distributions from UV and IR data, possibly demonstrating both that the UV SFRFs are (slightly) underestimated while IR SFRFs are overestimated. We perform the comparison of the SDSS DR7 SFRFs with the Illustris and IllustrisTNG simulations in section~\ref{TNGSFRF}.

The decline in the number density of galaxies below ${\rm SFR = } 10^{-1.5}$ ${\rm M_{\odot}yr^{-1}}$ shown in Fig. \ref{Fig1} is associated with the fact that the survey is incomplete and unable to detect numerous faint/low star-forming objects. The decline of the SDSS SFRF below this limit is not a behavior driven from physical reasons since the UV constrains given by \citet{Katsianis2017} which probe the SFRF to up to $10^{-2}$ ${\rm M_{\odot}yr^{-1}}$ and predictions from Cosmological simulations like EAGLE do not show this behavior and demonstrate a Schechter Form. Thus, we set as our confidence limit in SDSS in terms of galaxy SFRs at $10^{-1.5}$ ${\rm M_{\odot}yr^{-1}}$. The limit of SDSS in terms of the stellar mass is set at $10^9$ ${\rm M_{\odot}}$     \citep{weigel2016}.

\begin{table}
\bc
\begin{minipage}[htb]{140mm}
\caption[]{The intrinsic star formation rate function of SDSS DR7: the first column is the SFRs, and the second (third) column represents the corresponding comoving galaxy number densities (errors). The error bars are obtained by $150$ jackknife samples.
}
\label{tablel}\end{minipage}
\setlength{\tabcolsep}{10mm}
\small
 \begin{tabular}{ccccccccccccc}
  \hline\noalign{\smallskip}
$\log$ SFR & comoving galaxy number density & Error\\
${\rm [M_{\odot}yr^{-1}]}$ & ${\rm [dex^{-1}Mpc^{-3}]}$ & ${\rm [dex^{-1}Mpc^{-3}]}$ \\
 \hline\noalign{\smallskip}
$-2.85 $&$ 5.82\times10^{-4}$&$ 1.71\times10^{-4}$\\
$-2.55 $&$ 9.48\times10^{-4} $&$ 2.16\times10^{-4}$\\
$-2.25 $&$ 2.66\times10^{-3} $&$ 4.10\times10^{-4}$\\
$-1.95 $&$ 5.69\times10^{-3} $&$ 6.70\times10^{-4}$\\
$-1.65 $&$ 1.27\times10^{-2} $&$ 9.34\times10^{-4}$\\
$-1.35 $&$ 1.85\times10^{-2} $&$ 9.02\times10^{-4}$\\
$-1.05 $&$  1.98\times10^{-2} $&$ 8.23\times10^{-4}$\\
$-0.75 $&$ 1.68\times10^{-2} $&$ 5.20\times10^{-4}$\\
$-0.45 $&$  1.18\times10^{-2} $&$ 3.11\times10^{-4}$\\
$-0.15 $&$ 8.32\times10^{-3} $&$ 2.07\times10^{-4}$\\
$0.10 $&$ 5.41\times10^{-3} $&$ 1.21\times10^{-4}$\\
$0.33 $&$ 2.78\times10^{-3} $&$ 6.34\times10^{-5}$\\
$0.62 $&$  1.20\times10^{-3} $&$ 4.40\times10^{-5}$\\
$0.87 $&$ 3.66\times10^{-4}$&$  1.81\times10^{-5}$\\
$1.10 $&$ 8.33\times10^{-5} $&$ 7.82\times10^{-6}$\\
$1.26 $&$ 1.51\times10^{-5} $&$ 4.60\times10^{-6}$\\
$1.27 $&$ 2.19\times10^{-6} $&$ 2.41\times10^{-6}$\\
$1.48 $&$ 5.01\times10^{-7} $&$ 1.59\times10^{-6}$\\
$1.93 $&$ 1.98\times10^{-7}$&$  3.45\times10^{-7}$\\
$1.97 $&$ 3.81\times10^{-8} $&$ 1.96\times10^{-7}$\\
  \noalign{\smallskip}\hline
\end{tabular}
\ec
\end{table}

\subsection{The Intrinsic Specific Star Formation Rate Function}
\label{sub:ssfrf}
A direct measurement of the connection between galaxy star formation rates and stellar masses involves the specific star formation rate \citep{Brinchmann00}, defined as the SFR per unit stellar mass ${\rm M^*}$, i.e., ${\rm sSFR = SFR/M^*}$. The specific Star Formation Rate (sSFR) of a galaxy is a key property commonly used in the literature to distinguish if the galaxy is star-forming or quenched. It is a common practice to define the passive population as galaxy with sSFR lower than ${\rm 10^{-11} \, yr^{-1}}$ \citep{Ilbert2015,Katsianis2020}. Thus, constructing the Specific Star Formation Rate Function (sSFRF) enables us to study quantitative and qualitative the distribution of the quenched and star-forming objects in SDSS DR7 and simulations.

\begin{figure} 
   \centering
   \includegraphics[width=0.7\textwidth]{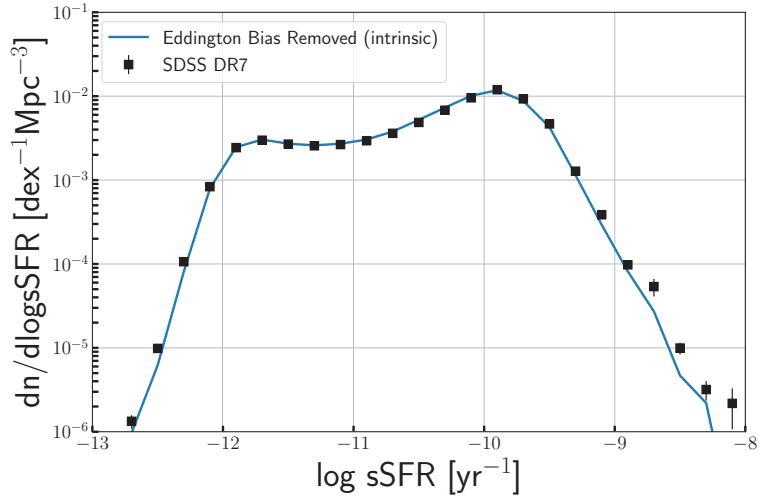}
   \caption{The Eddington bias correction on specific star formation rate 
   in SDSS DR7. The black dots represent the observed 
   sSFRF while the blue line is the Eddington bias corrected sSFRF. 
   The error bars on the observed sSFRF are computed from 150 jackknife samples.
   } 
   \label{SSFR_EB}
   \end{figure}

Following the steps laid out in 
section \ref{subsec:EB_correction}, 
one could have applied the methodology to the observed sSFRF. 
However, the shape of the sSFRF 
(bi-modal form, shown in Fig.~\ref{SSFR_EB}) 
raises the difficulty of applying our method, 
which works only for the Schechter-like function. 

Instead, by utilizing the by-products of SFRF Eddington
bias correction procedures, i.e., the approximated
Eddington bias corrected SFRs for the individual galaxy, 
one would immediately have the Eddington-bias-free sSFRs 
for each galaxy once the stellar mass is corrected for the Eddington bias as well. 
Luckily the stellar mass function (SMF) also follows a Schechter function shape and 
it allows us to apply our method to the stellar mass function so that 
we could obtain approximated
Eddington-bias-corrected stellar masses for each galaxy

We correct the Eddington bias in the stellar mass function in 
appendix~\ref{subsec:SMF_EB_appendix} (see the Fig.\ref{fig:SMF_EB}).
Note that our method is robust in terms of recovering
the intrinsic SFRF/SMF, as proven in section \ref{subsec:EB_correction}.
However, it is not necessarily exact in extracting the correction
 down to the level of individual galaxies. As an approximation, we support 
that it is a valid approach to do Eddington bias corrections for the case of sSFRF.

The inferred Eddington-bias free sSFRF is shown in Fig.~\ref{SSFR_EB},
which is almost identical to that without Eddington bias correction, 
except at the very active star-forming regime (${\rm  sSFR \sim 10^{-8} yr^{-1}}$). 
The similarity between the two sSFRs is probably due to the cancellation of the Eddington bias in both SFRF and stellar mass function.
We limit our analysis to galaxies with $ {\rm SFR  > 10^{-1.5} \, M_{\odot} yr^{-1}}$  and stellar mass $ {\rm M_{\star} > 10^9 \, M_{\odot}}$ for completeness reason.

\section{Comparison with cosmological simulations}
\label{sect:Comp}

\subsection{The IllustrisTNG simulations}

Illustris-1 \citep{Vogelsberger2014} consists a cosmological simulation run with the moving-mesh code AREPO \citep{springel2010}. It includes sophisticated sub-grid physics that involve gas cooling, sub-resolution inter-stellar medium modeling, stochastic star formation, stellar evolution, gas recycling, chemical enrichment,  kinetic stellar feedback driven by SNe explosions, supermassive black hole (SMBH) growth and related AGN feedback. The IllustrisTNG \citep{2017MNRAS.465.3291W,Pillepich2018} project is the successor of the Illustris simulations and includes an updated galaxy formation model that employs new physics and numerical improvements to address some shortcomings of the original Illustris-1 model \citep{Pillepich2018}. Some key and notable improvements relevant to our work are :
\begin{itemize}
\item  An updated kinetic AGN feedback model for objects with low accretion rates in the form of  a  kinetic,  super-massive driven wind \citep{2017MNRAS.465.3291W}.  The above implementation enhances feedback, especially for objects with $10^{12}-10^{14} \, M_{\odot}$  halo masses, and decreases the simulated stellar masses for the TNG100 model bringing observed and simulated stellar mass functions in better agreement \citep{Pillepich2018}. In contrast, Illustris reproduced a stellar mass function with higher values at z $<$ 1.
\item An improved parameterization of galactic winds \citep{Pillepich2018}. Differently from Illustris, winds are injected isotropically, with larger wind Velocity and Energy Factors. The new feedback implementation solved the mild decline in the cosmic star formation rate density of the original Illustris model at z $<$1 and played a major role in shaping the stellar mass function of low mass objects with ${\rm M_{\star} < 10^{10} \, M_{\odot}}$.
\item The updated TNG model produces the observed color bi-modality. \citet{Nelson2017} demonstrated that the  simulated ${\rm (g-r)}$ colors of TNG galaxies at low redshift are in good agreement with a quantitative comparison to observational data from SDSS at $z<0.1$. The authors obtained the locations in the color of both the red and blue populations at the ${\rm color-M_{\star}}$ plane, the relative strength between the red and blue distributions using histograms of ${\rm (g-r)}$ colors, the location of the color minimum between the two populations, and the location of the maximal point of the bi-modality. The authors suggested that this is the result of the updated feedback prescriptions in the improved next-generation model.

\end{itemize}
The TNG300-1 and TNG100-2 simulations are performed at a factor of $8$ lower in mass and $2$ at spatial resolution when compared to the TNG100 run. Otherwise all 3 configurations adopt an identical model with the same parameters for their sub-grid models regardless of box-size and resolution. TNG100 has a similar resolution as the original Illustris simulation so we can perform a direct comparison between them. More details for the simulations are summarized in table \ref{simulation}.

\subsection{IllustrisTNG star formation rate function and Cosmic star formation rate density}
\label{TNGSFRF}

In Fig. \ref{Fig1} we demonstrate that the Illustris star formation rate function has a higher normalization with respect to our SDSS observations at all SFR regimes, while the TNG100-1 simulation performs much better, especially for objects with low SFRs. The reason for this is that the updated TNG model includes a range of improvements (e.g. on the AGN feedback and galactic winds schemes) to decrease the simulated stellar masses and cosmic star formation rate density at $z < 1$. We demonstrate that the TNG100-1 star formation rate function has a good agreement with the SDSS observations for objects with ${\rm SFR = 0.01 - 5 \, M_{\odot}yr^{-1}}$. 
However, the TNG model does not reproduce our SDSS observations at the high star-forming end and typically lay between the UV and IR constraints given in \citet{Katsianis2017}. In other words, TNG300-1 reproduces the observed SDSS SFRF at the  ${\rm SFR > 5 \, M_{\odot}  yr^{-1}}$ regime. It would be intriguing to suggest that this agreement happens since the TNG300-1 simulation has a larger box-size and thus can sample a larger number of objects, employ better statistics and is more trust-worthy at the high star-forming end, making the effects of finite box-size less severe. However, this good agreement between TNG300-1 and SDSS at the high star-forming end is a matter of coincidence and an effect of low resolution (more details can be found in the Appendix. \ref{appen.reso_box_test}). In the Appendix.\ref{appen.csfrd}, we also calculate the cosmic star formation rate density in the local universe since it is a cosmic metric for star formation usually employed in the literature and perform further resolution and box-size tests.

\begin{table}
\bc
\begin{minipage}[]{140mm}
\caption[]{Primary parameters of the simulations analysed in this study. column1, run name; column2, box volume of the simulation; column3, number of dark matter particles; column4, mass of the dark matter particles; column5, initial mass of the gas particles; column6, number of galaxies at $z = 0$
\label{simulation}}\end{minipage}
\setlength{\tabcolsep}{7.5pt}
\small
 \begin{tabular}{ccccccccccccc}
  \hline\noalign{\smallskip}
Simulation Name & Volume(Mpc$^3$)  & ${\rm N_{DM}}$ & ${\rm m_{DM}(10^6M_{\odot})}$ & ${\rm m_{gas}(10^6M_{\odot})}$ & ${\rm N_{galaxy}}$$(z=0)$ \\
 \hline\noalign{\smallskip}
Illustris-1 &$106.5^3$ & $1820^3$ & 6.3 & $1.3$ & $4366546$ \\
TNG100-1 & $110.7^3$ & $1820^3$ & 7.5 & $1.4$ & $4371211$ \\
TNG100-2 & $110.7^3$ & $910^3$ &  60 & $11$ & $698336$ \\
TNG100-3 & $110.7^3$ & $455^3$ &  480& $89$ & $118820$ \\
TNG300-1 & $302.6^3$ & $2500^3$ & 59& $11$ & $14485709$\\
  \noalign{\smallskip}\hline
\end{tabular}
\ec
\end{table}

\subsection{The IllustrisTNG specific star formation rate function} 
\label{TNGsSFRF}

\begin{figure} 
   \centering
   \includegraphics[width=0.7\textwidth]{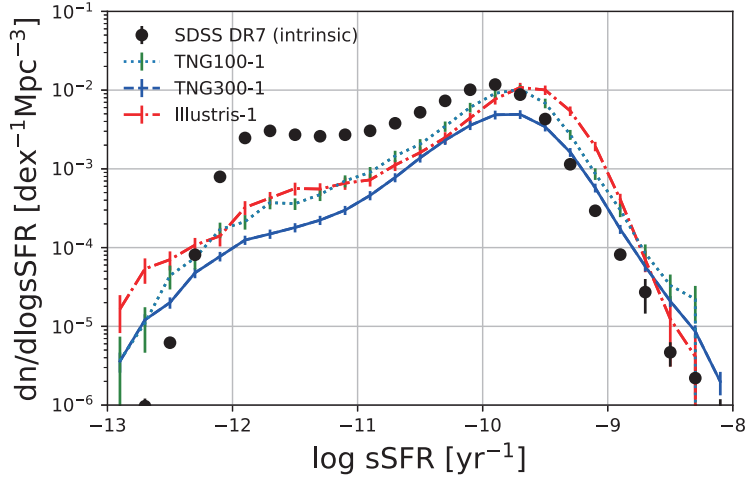}
   \caption{The specific star formation rate of simulations and observation: the green dashed line represents the sSFRF derived from TNG100-1, the blue solid line from TNG300-1, and the red dash-dotted line from Illustris-1. The black filled dots are intrinsic sSFRF inferred from SDSS DR7, with correction for the Eddington bias (see subsection \ref{sub:ssfrf} for details), the error bars are obtained by the Jacknife method.} 
   \label{SSFR}
   \end{figure}

The fact that TNG-100 can reproduce consistent star formation rate functions and stellar mass functions with SDSS does not necessarily mean that this is achieved with simulated galaxies that each uniquely fulfills the observed relation between SFR and stellar mass. We investigate how the simulated Specific Star Formation Rate Function (sSFRF) from the Illustris and IllustrisTNG simulations compares with SDSS in Fig. \ref{SSFR}. We impose the same limits to the simulations ($ {\rm SFR  > 10^{-1.5} \, M_{\odot} yr^{-1}}$  and stellar mass $ {\rm M_{\star} > 10^9 \, M_{\odot}}$). We see that both the observed and simulated distributions have a peak at ${\rm sSFR \sim 10^{-9.7} yr^{-1}}$. These galaxies would be classified as star-forming objects and it is encouraging that TNG can reproduce qualitative this behavior. We note that this is found by the model regardless of resolution (more details in the Appendix. \ref{appen.reso_box_test}) 

\begin{table}
\bc
\begin{minipage}[]{130mm}
\caption[]{specific star formation rate of SDSS, the first column is the median value of the interval, the second column is corresponding number of galaxies, and the last column is the error calculated by Jacknife method.}
\label{tab:SSFR}\end{minipage}
\setlength{\tabcolsep}{10mm}
\small
 \begin{tabular}{ccccccccccccc}
  \hline\noalign{\smallskip}
$\log$ sSFR & comving galaxy number density & Jacknife error\\
 ${\rm [yr^{-1}]}$ &[${\rm dex^{-1}Mpc^{-3}}$]&[${\rm dex^{-1}Mpc^{-3}}$]\\
 \hline\noalign{\smallskip}
 $-12.9$ &$4.32\times10^{-7}$ & $1.90\times10^{-7}$\\
 $-12.7$ & $9.68\times10^{-7}$ & $2.42\times10^{-7}$\\
 $-12.5$ & $6.20\times10^{-6}$ &$6.41\times10^{-7}$\\
 $-12.3$ & $8.11\times10^{-5}$ &$4.99\times10^{-6}$\\
 $-12.1$&$7.92\times10^{-4}$ &$2.24\times10^{-5}$\\
 $-11.9 $&$2.47\times10^{-3}$& $6.50\times10^{-5}$\\
 $-11.7$&$3.04\times10^{-3}$ & $8.00\times10^{-5}$\\
 $-11.5 $&$2.71\times10^{-3}$ & $7.17\times10^{-5}$\\
$ -11.3 $&$2.59\times10^{-3}$ & $8.02\times10^{-5}$\\
$ -11.1 $&$2.71\times10^{-3}$ &$9.58\times10^{-5}$\\
$ -10.9 $&$3.04\times10^{-3}$ &$1.22\times10^{-4}$\\
 $-10.7$&$3.79\times10^{-3}$ &$1.55\times10^{-4}$\\
$ -10.5 $&$5.23\times10^{-3}$ &$1.72\times10^{-4}$\\
 $-10.3 $&$7.31\times10^{-2}$ &$1.97\times10^{-4}$\\
 $-10.1 $&$1.01\times10^{-2}$ &$2.58\times10^{-4}$\\
 $-9.9  $&$1.18\times10^{-2}$ &$3.09\times10^{-4}$\\
 $-9.7  $&$8.76\times10^{-3}$ &$2.58\times10^{-4}$\\
 $-9.5  $&$4.30\times10^{-3}$ &$1.32\times10^{-4}$\\
 $-9.3  $&$1.14\times10^{-3}$ &$4.34\times10^{-5}$\\
 $-9.1  $&$2.94\times10^{-4}$ &$3.60\times10^{-5}$\\
 $-8.9  $&$8.15\times10^{-5}$ &$8.58\times10^{-6}$\\
 $-8.7  $&$2.72\times10^{-5}$ &$1.27\times10^{-5}$\\
 $-8.5  $&$4.68\times10^{-6}$ &$1.60\times10^{-6}$\\
 $-8.3$&$2.20\times10^{-6}$ &$8.32\times10^{-7}$\\
 $-8.1$&$8.01\times10^{-8}$ &$1.11\times10^{-6}$\\
  \noalign{\smallskip}\hline
\end{tabular}
\ec

\end{table}

The intrinsic specific star formation rate function of SDSS DR7 shown in Fig. \ref{SSFR} demonstrates a clear bi-modality. To be more specific a second peak is detected at ${\rm sSFR \sim 10^{-12} yr^{-1}}$ reflecting the presence of the quenched population of galaxies, which at redshift $z \sim 0$ is expected to be abundant. We note that this population does not appear in IllustrisTNG which does not exhibit the same qualitative behavior with a double peak. Quantitative the TNG run has almost an order of magnitude lower number density of objects with ${\rm sSFR \sim 10^{-12} yr^{-1}}$ comparing with the SDSS constraints. The reason for this tension possibly reflects the need for inclusion of a different or more effective quenching mechanism.  We note that there are no significant deviations of the sSFRFs of the TNG model from the original Illustris simulation.



\section{Conclusions}
\label{sect:Res}

In this work we present the first Eddington-bias-free star formation rate function (SFRF), cosmic star formation rate density (CSFRD) and specific star formation rate function (sSFRF) in the Sloan Digital Sky Survey Data Release 7. We perform comparisons of the above observational constraints with the reference simulations of Illustris and IllustrisTNG. We include resolution tests and discuss the accomplishments and shortcomings of the models.
In the following we summarize the main results and conclusions of our analysis:
\begin{itemize}

 \item Without resort to assuming a function form 
for the intrinsic (Eddington bias corrected) SFRF, we correct the Eddington bias on the SFRF and sSFRF by subtracting the SFR of each galaxy using the average shift in the SFRF 
induced by the Eddington bias iteratively (Fig. \ref{fig:EBcorrection}). 
We test our method on a simulated Eddington biased
SFRF from TNG100-1 and the inferred ``intrinsic'' SFRF matches well with true
SFRF (Fig. \ref{fig:simulation_proof}). The test reflects the
robustness of our method and in principle, it could be generalized 
to any Schechter-like function. We apply the above method to the SDSS SFRF and compare our results with predictions from cosmological simulations and other SFR indicators.\\

\item The SFRF constructed from the SED derived star formation rates of the SDSS survey are in excellent agreement with the SFRFs obtained from UV luminosities for objects at the $ {\rm SFR \sim 0.01 - 5 \, M_{\odot} yr^{-1}} $ regime presented in \citet{Katsianis2017}. However, the  high star-forming end ($ {\rm SFR > 10 \, M_{\odot} yr^{-1}} $) lays between the determinations of the UV and IR/radio tracers. For the high star formation rates a tension between UV and IR indicators is established in the literature owning to either underestimations of UV SFRs or over-estimations of the IR SFRs. The SDSS SED star formation rate function of this work is in good agreement with other star formation rate indicators, especially UV which is able to probe low star forming objects, up to $ {\rm SFR = 10^{-1.5} \, M_{\odot} yr^{-1}} $ . Thus we set our confidence limit for SFRs to this value. \\

\item The simulated reference model of the IllustrisTNG labeled as TNG100-1 produces a SFRF that is consistent with the constraints of the SDSS data for objects at the $ {\rm SFR \sim 0.01 - 5 \, M_{\odot}yr^{-1}} $ regime, while it performs much better than the original Illustris model. This reflects the improvements taken into account in the updated TNG model, including the feedback prescriptions. However, the simulation does not perform equally good for higher star-forming objects ($ {\rm SFR > 10 \, M_{\odot} yr^{-1}} $ ) with observations having lower number densities. The configuration with 8 times lower resolution and $ \sim $20 times larger volume (labeled as TNG300-1) demonstrates a better agreement at the high star-forming end, despite the fact it is not as successful for low star-forming objects. However, the reason for this agreement is coincidental and has its roots in resolution effects, rather than the better statistics produced in the larger box-size (appendix \ref{appen.reso_box_test}). This resolution driven effect brings observed and simulated high star-forming ends in agreement for no physical motivated reasons.\\

\item We demonstrate that the intrinsic specific star formation rate function from SDSS has two peaks and demonstrates a clear bi-modality for objects with  $ {\rm SFR  > 10^{-1.5} \, M_{\odot} yr^{-1}}$  and stellar mass $ {\rm M_{\star} > 10^9 \, M_{\odot}}$. The one peak appears at ${\rm  sSFR \sim 10^{-9.7} yr^{-1}}$. These galaxies would be classified as star-forming objects. A second peak is detected at ${\rm  sSFR \sim 10^{-12} yr^{-1}}$ reflecting the presence of the quenched population of galaxies, which at redshift $z \sim 0$ is expected to be abundant (subsection \ref{TNGsSFRF}). \\

\item  We note that the bi-modal, two peaked sSFRF implied by SDSS observations does not appear in TNG100-1 or TNG300-1. The simulations do not exhibit the same qualitative behavior and demonstrate only one peak for high star-forming objects at ${\rm  sSFR \sim 10^{-9.6} yr^{-1}}$. The TNG run has almost 1 order of magnitude lower number density of passive objects with ${\rm sSFR \sim 10^{-12} yr^{-1}}$ with respect observations. This tension may reflect the need for inclusion of an additional or more efficient quenching mechanism  (subsection \ref{TNGsSFRF}). We note that the normalization of the simulated sSFRF increases with resolution but its shape remains the same.\\

\end{itemize}

\normalem
\begin{acknowledgements}
The authors thank the anonymous referee for the helpful comments that significantly improve the presentation of this paper.
HX thanks the useful discussion with Jiajun Zhang and Zhaozhou Li.
This work is supported by the national science foundation of China 
(Nos. 11833005,  11890692, 11621303), 111 project No. B20019 and
Shanghai Natural Science Foundation, grant No. 15ZR1446700.
We gratefully acknowledge the support of the Key Laboratory for Particle Physics, Astrophysics and Cosmology, Ministry of Education and the Tsung-Dao Lee Institute.
\end{acknowledgements}

\bibliographystyle{raa}
\bibliography{ms2020-0141}

\appendix

\section{A test of the Eddinghton Bias correction method using TNG100}
\label{appendixilleb}

\begin{figure} 
   \centering
   \includegraphics[width=0.7\textwidth]{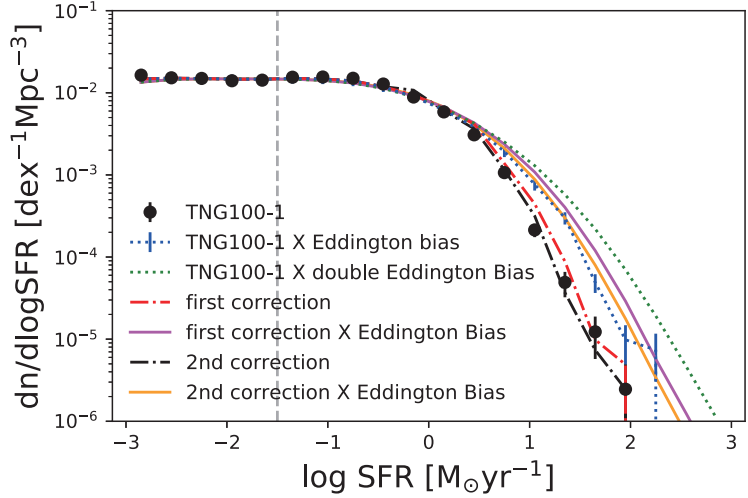}
   \caption{A test of our Eddington bias correction method on the star formation rate function using TNG100-1. The black dots represent intrinsic values from TNG100-1. The blue (green) dashed line represents intrinsic values plus (twice) Eddington bias while the red (blue) dash-dotted line is the first (second) correction. And the magenta(orange) solid line is the first (second) correction plus Eddington bias. Based on our criteria in Step 5 of section ~\ref{subsec:EB_correction}, the second correction is ought to be the ``intrinsic SFRF'' and it matches well with that from the simulation. 
   The error bars shown on the SFRFs are computed 
   from $64$ jackknife samples.
   To mimic the observation, the Eddington biased
   SFRF is computed from {\it only one} mock, which induces some
   wiggles in the curves. The twice Eddington biased
   and corrections convolved with Eddington
   bias SFRF are estimated from 1000 mocks.
   See details in section ~\ref{subsec:EB_correction}.
   }
   \label{fig:simulation_proof}
   \end{figure}

To ensure that our method recovers (or at least approaches as close as possible to) 
the intrinsic SFRF, 
we test our method in the TNG100-1 simulation in Appendix \ref{appendixilleb}. 
The first step is to start with an Eddington-biased SFRF, 
which is the counterpart to the observed SFRF in the simulation. 
However, due to the lack of Eddington bias in simulation, 
we have to manually assign some uncertainties in SFR 
for all the simulated galaxies to mimic the observed SFRF. 
The observation suggests that the error in SFR should be dependent on SFR,
but for simplicity, we assume a universal 0.4 dex uncertainty on their SFRs
 (the arithmetic mean for all the SDSS DR7 galaxies error 
provided in the MPA-JHU catalog) for all galaxies. 
We note that assigning an SFR-dependent error in SFR would not 
change the main conclusion in this test. 
Given the simulated Eddington-biased SFRF and 
the assumed universal SFR uncertainties,
we obtain the ``intrinsic SFRF'' by following 
the steps outlined in section \ref{subsec:EB_correction}, which 
turns out be an excellent match with the true SFRF 
directly from the simulation, shown in Fig. \ref{fig:simulation_proof}.

\section{The Eddington Bias Corrected Stellar Mass Function}
\label{subsec:SMF_EB_appendix}

\begin{figure} 
   \centering
   \includegraphics[width=0.7\textwidth]{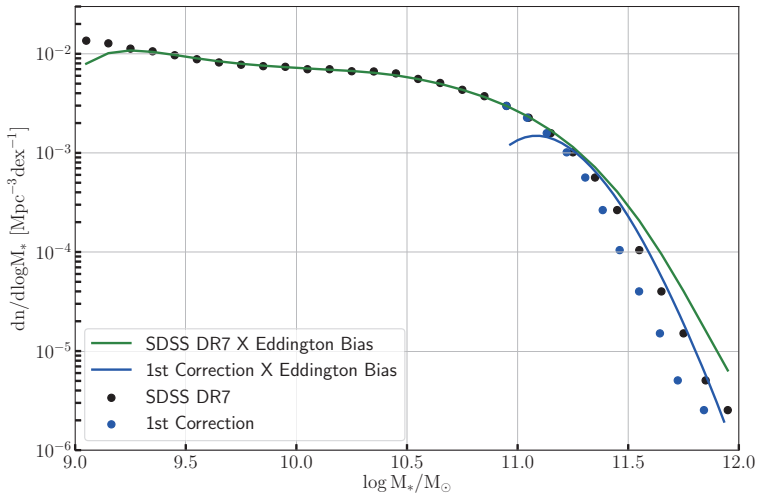}
   \caption{The Eddington bias correction on the stellar mass function for SDSS DR7. The black dots represent the observed SFRF directly from SDSS DR7. The green (blue) solid line represents the SDSS DR7 (first correction) plus Eddington bias. The blue dots represent the first correction and the intrinsic SMF we are seeking. The flattening behavior on the left side of blue solid line is
   totally artificial because we only build histograms with updated galaxy SFRs, same as in Fig. \ref{fig:EBcorrection}. 
   See more details in  section \ref{subsec:EB_correction} and Appendix.\ref{subsec:SMF_EB_appendix}.
   } 
\label{fig:SMF_EB}
\end{figure}

Following the steps laid out in section~\ref{subsec:EB_correction}, 
we start with the stellar mass provided by the JHU-MPA group. 
It only takes 1 iteration to obtain the intrinsic stellar mass function (see
Fig.~\ref{fig:SMF_EB}). We note that a universal stellar mass error, $\sim$ 0.15 dex \citep{Li2009,Yang2012}, is assumed during the procedures. 

\section{resolution test and box-size test}
\label{appen.reso_box_test}
\begin{figure}[h]
\hspace{1.0cm}
\begin{subfigure}{0.45\textwidth}
\includegraphics[width=0.95\linewidth, height=5.5cm]{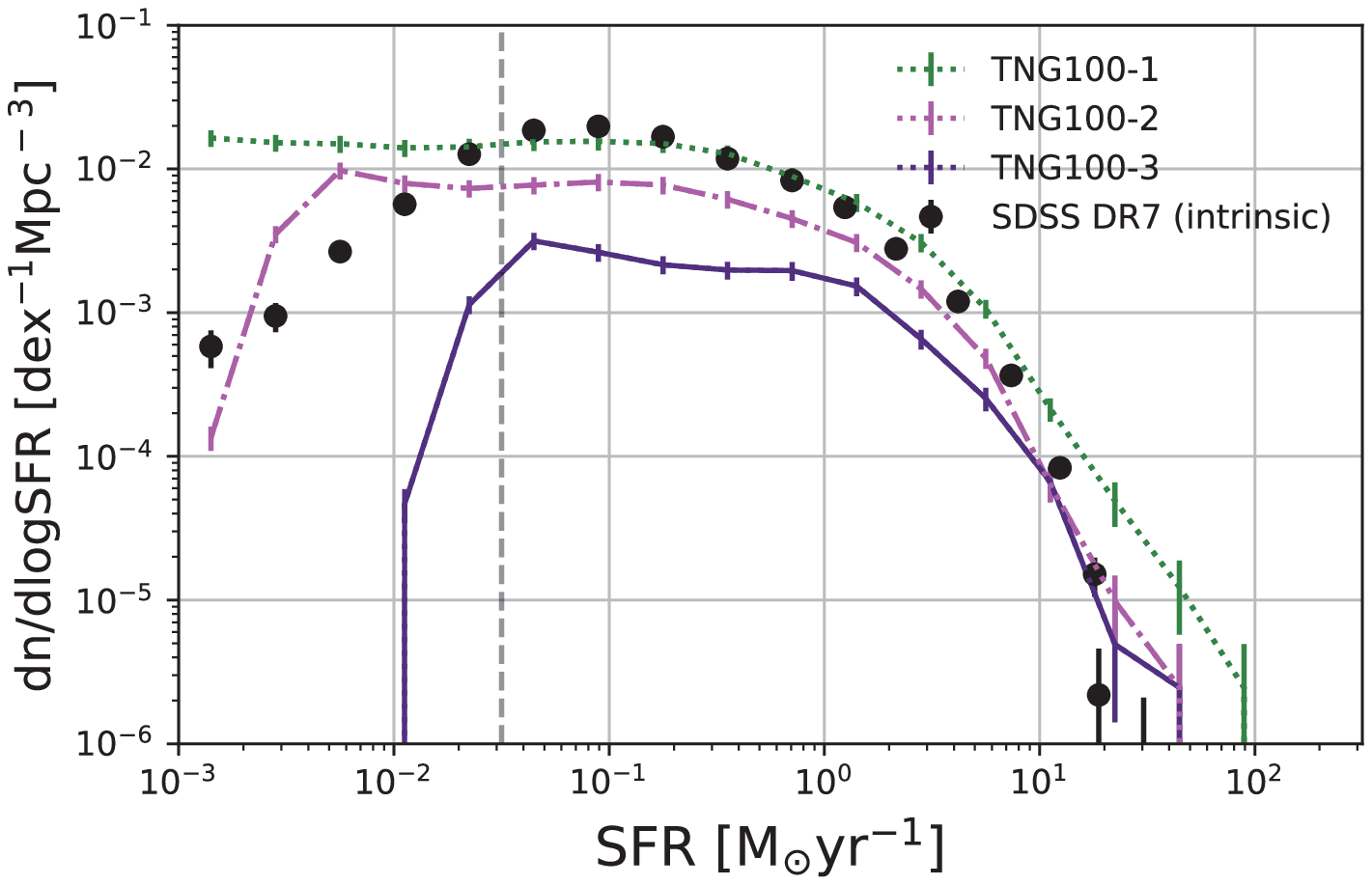}
\caption{resolution test}
\label{fig:total_compare1}
\end{subfigure}
\begin{subfigure}{0.45\textwidth}
\includegraphics[width=0.95\linewidth, height=5.5cm]{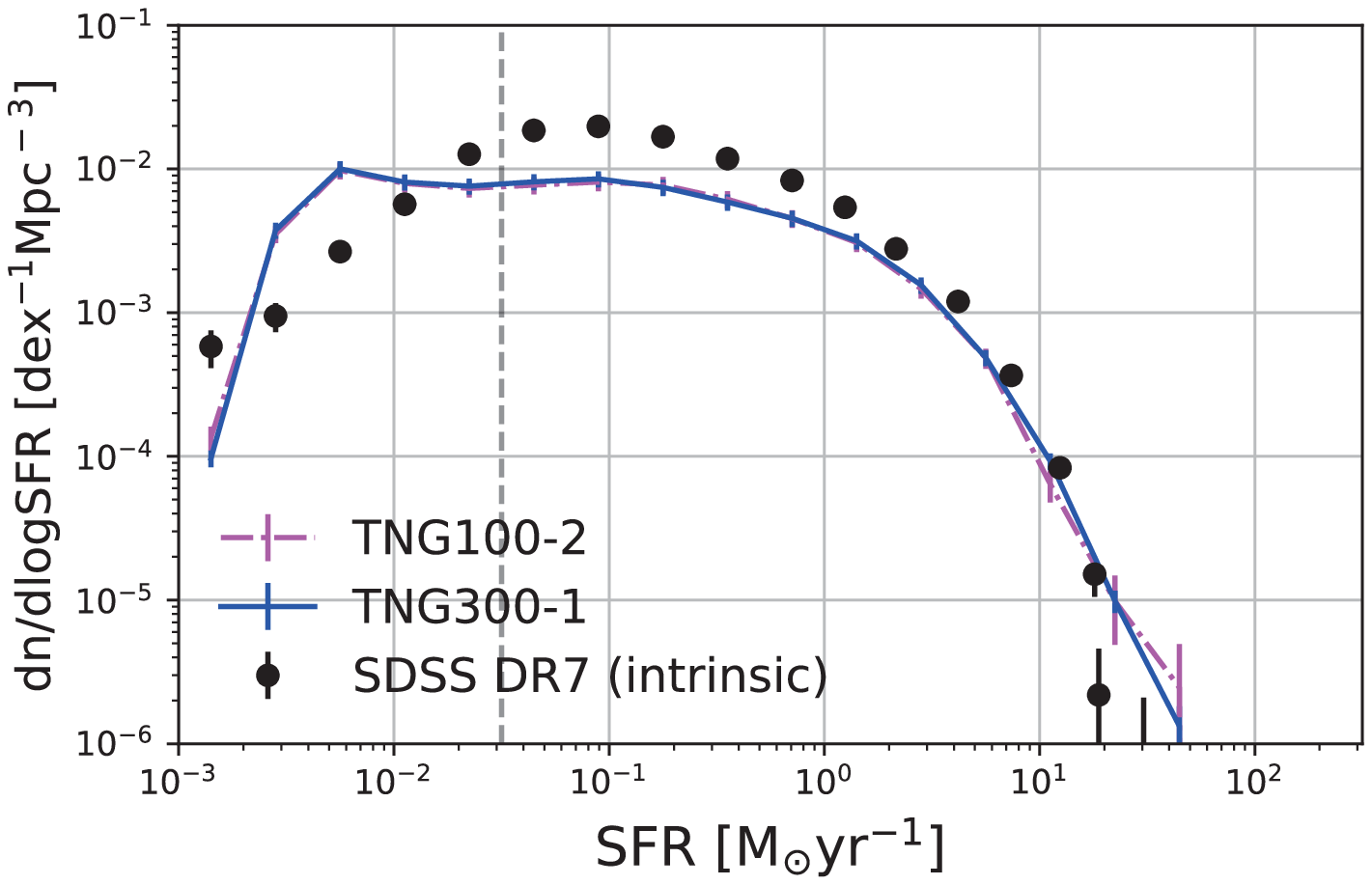}
\caption{box-size test}
\label{fig:total_compare2}
\end{subfigure}
\caption{The SFRF between TNG simulations with different resolution and box-sizes. a) left panel : This is a test that shows how the model varies with resolution as these $3$ simulations have the same box-size but different particle masses. Green dotted line represents TNG100-1, the magenta dash-dotted line represents TNG100-2, the indigo solid line represents TNG100-3, and the black dots represent SDSS DR7. b) Right panel : This is a test which shows how the model varies with box-size as these $2$ simulations have the same resolution but different volume. The blue solid line represents TNG300-1, the error bars in plots are obtained by Jacknife method.}
\label{fig:image2}
\end{figure}

At the left panel of Fig. \ref{fig:image2} the comparison between TNG100-1 and TNG100-2 demonstrates that the simulated SFRF is highly dependent on resolution, even at the high star-forming end, and the TNG100-2 run has better agreement with the SDSS observations for objects with ${\rm SFR > 5 \, M_{\odot}  yr^{-1}}$. At the right panel of Fig. \ref{fig:image2} we demonstrate that the TNG300-1 and TNG100-2 simulations reproduce identical results (both have the same resolution which is 8 times lower in mass from TNG100-1). The perfect agreement between TNG300-1 and TNG100-2 possibly reflects that the box-size of 100 Mpc is enough for studies of the cosmic star formation rate density and SFRF in the TNG model. The complement of TNG100-1 simulation seems not affected significantly in terms of galaxy SFRs or low numbers of galaxies at the high star-forming end by the smaller box-size. However, the TNG100-1 and TNG100-2 simulations do not converge at any SFR regime pointing to limitations of the model related to resolution. \\

\begin{figure}[h]
\hspace{1.0cm}
\begin{subfigure}{0.45\textwidth}
\includegraphics[width=0.95\linewidth, height=5.5cm]{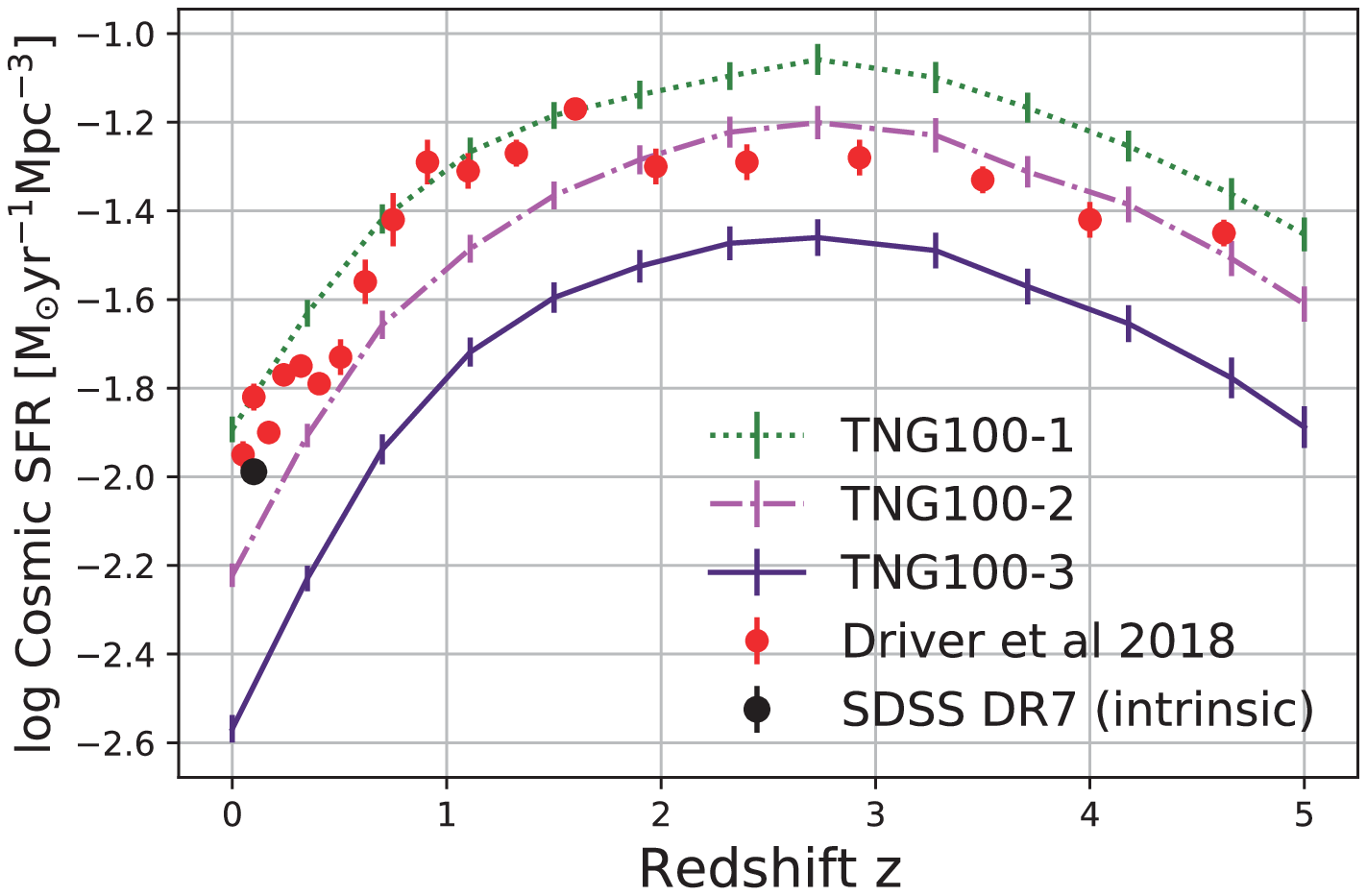} 
\caption{resolution test}
\label{fig:csfrd1}
\end{subfigure}
\begin{subfigure}{0.45\textwidth}
\includegraphics[width=0.95\linewidth, height=5.5cm]{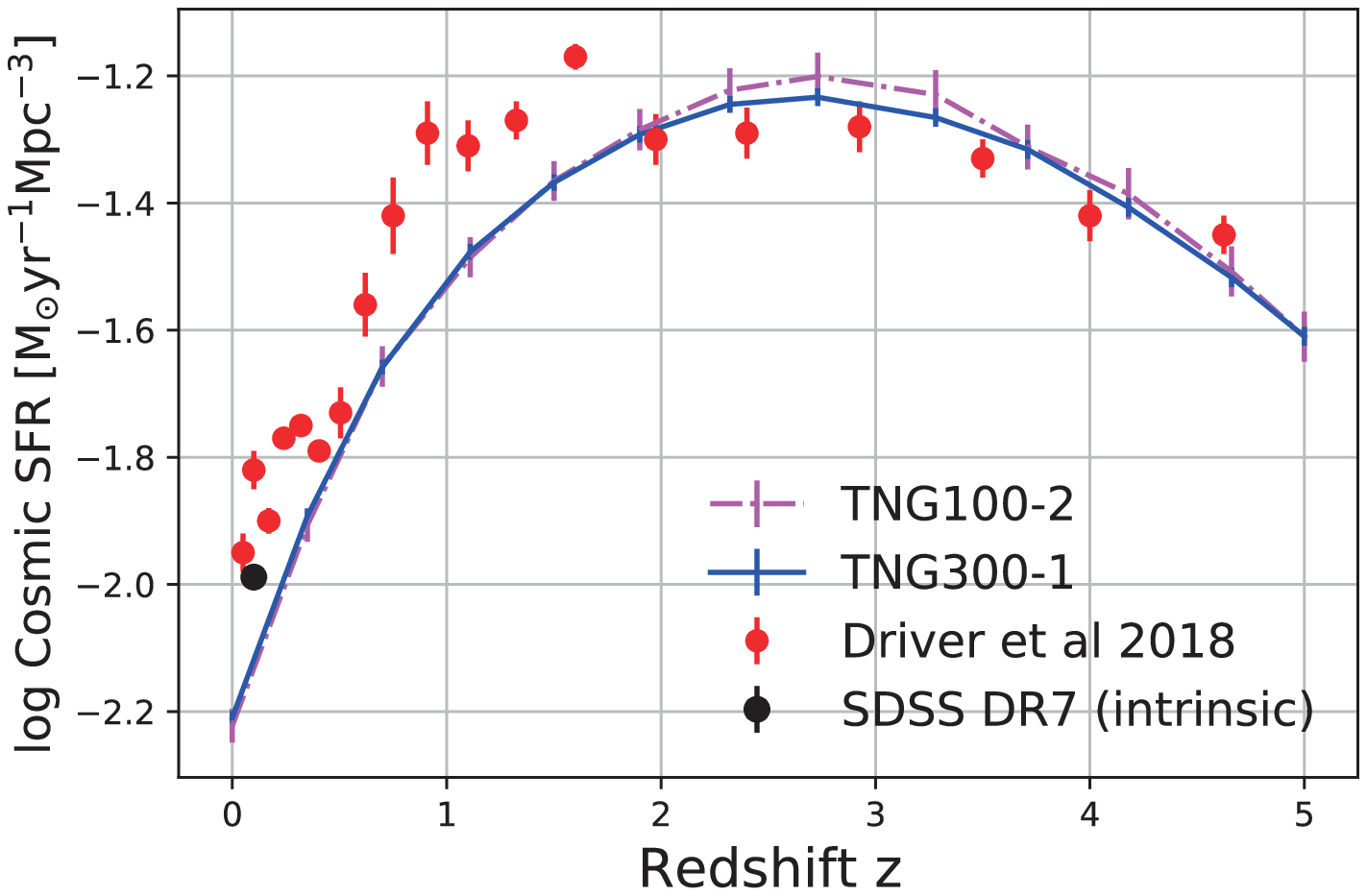}
\caption{box-size test}
\label{fig:csfrd2}
\end{subfigure}
\caption{Comparison of the Cosmic Star formation densities of different simulations. Similar to Fig.\ref{fig:image2}, the left panel and right panel represent the resolution test and box-size test, respectively. The green dotted line represents TNG100-1, the magenta dash-dotted line represents TNG100-2, the indigo solid line represents TNG100-3, the blue solid line represents TNG300-1, the black dot represents the intrinsic value of SDSS DR7, and magenta dots represent observation data from \citet{driver2018}. the error bars in the plots are obtained by Jacknife 
method.}
\label{fig:csfrd22}
\end{figure}

\begin{figure}[h]
\hspace{1.0cm}
\begin{subfigure}{0.45\textwidth}
\includegraphics[width=0.95\linewidth, height=5.5cm]{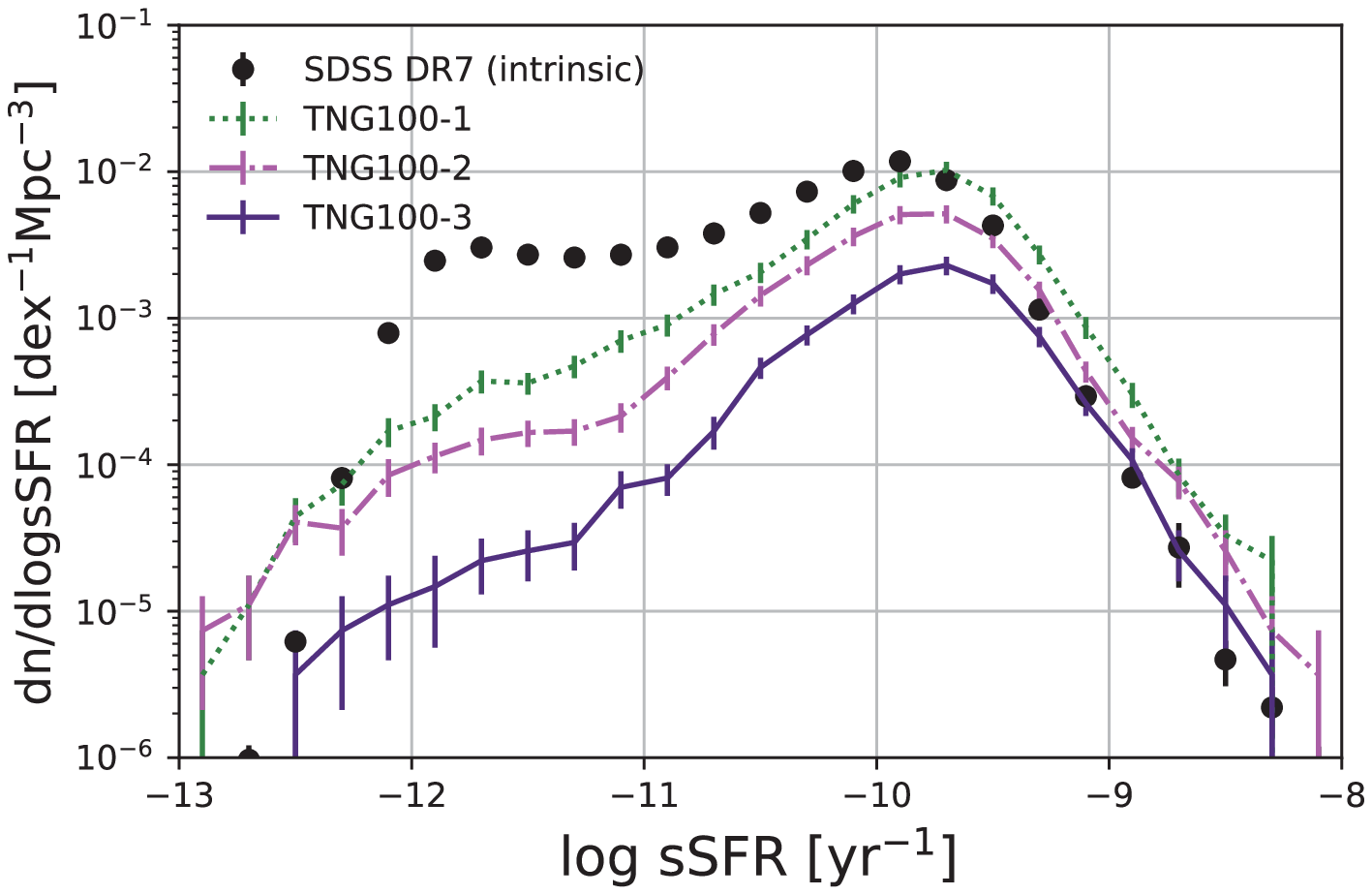} 
\caption{resolution test}
\end{subfigure}
\begin{subfigure}{0.45\textwidth}
\includegraphics[width=0.95\linewidth, height=5.5cm]{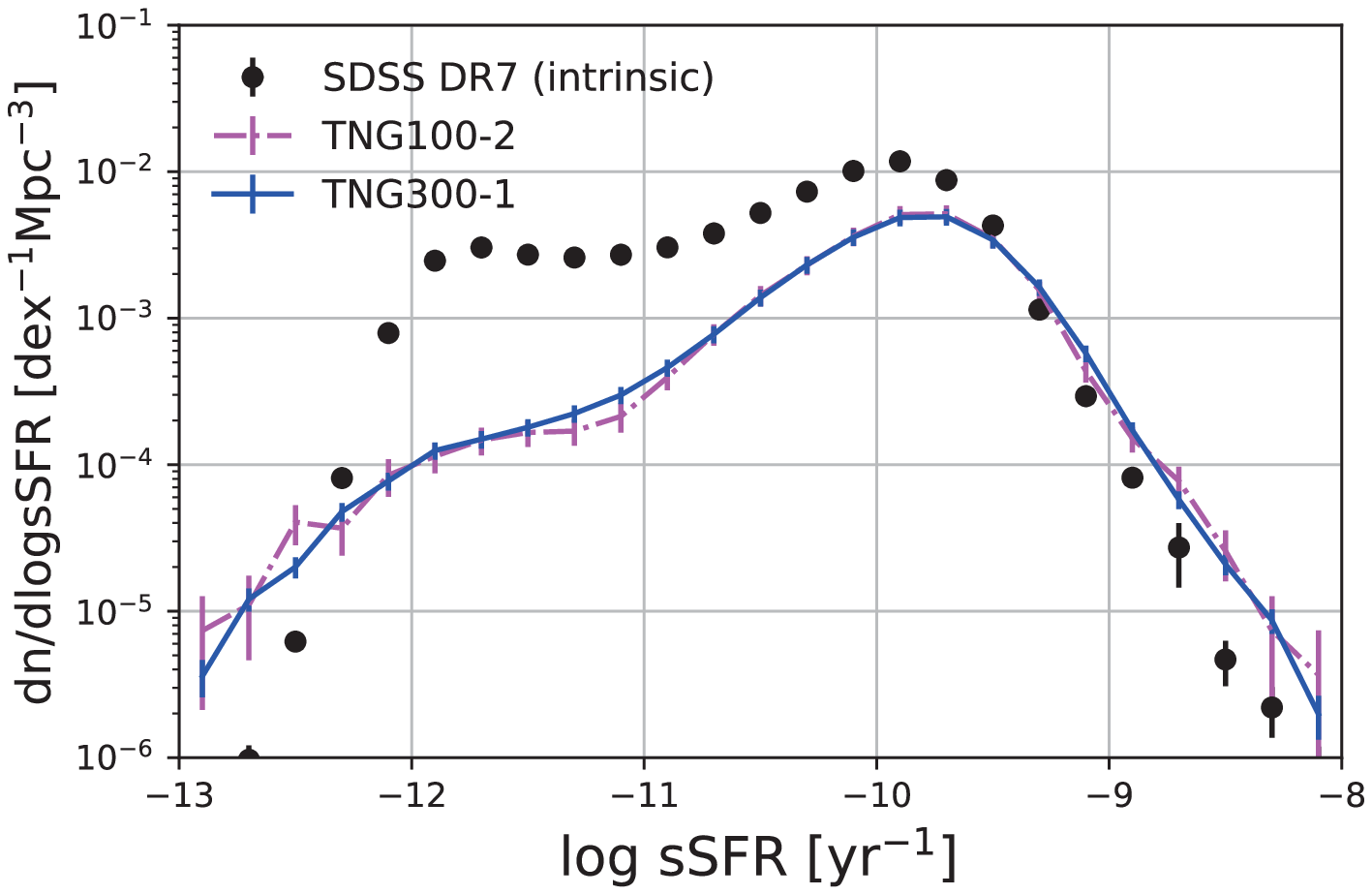}
\caption{box-size test}
\end{subfigure}
\caption{a)This is a resolution test: The green dotted line represents TNG100-1, the magenta dash-dotted line represents TNG100-2, the indigo solid line represents TNG100-3, and the black dots from SDSS DR7 are attached for reference, b)This is a box-size test, the blue solid line represents TNG300-1. The error bars in plots are obtained by the Jacknife method}
\label{fig:image22}
\end{figure}

We note that all TNG runs, regardless of resolution or box-size use the default model parameter values given in \citet{Pillepich2018} and no adjustments with resolution were done. \citet{2015MNRAS.446..521S} discussed the importance of re-scaling the parameters (especially feedback) of higher resolution simulations to produce properties and statistics of galaxies that converge with the lower resolution runs. The above convergence test (the agreement between the high resolution simulation with the one that adopts lower resolution and re-scaled parameters for sub-grid Physics) was labeled by the authors as the ``weak convergence" test, which EAGLE SFRFs satisfy \citep{Katsianis2017}. The ``strong convergence" test is only fulfilled when convergence between low and high resolution simulations is satisfied without any re-scaling of the parameters and consists of the ultimate test for the independence of the adopted cosmological model on the resolution. We demonstrate that the ``strong resolution convergence" is not satisfied for the Illustris TNG star formation rate functions and the higher resolution TNG100-1 run does not converge with the TNG100-2 and TNG300-1 runs, having a larger normalization by 2 times at all star formation rates regimes.  \citet{Pillepich2018b} demonstrated that TNG100-1 and TNG300-1 stellar mass functions would come into agreement by re-scaling the lower resolution simulation by a factor of 1.4. The authors emphasized that, while the incomplete resolution convergence of the stellar mass functions of TNG300-1 with the TNG100-1 is without a  doubt a limitation of the model the needed re-scaling factor of 1.4 is relatively small and comparable with the current discrepancies across different observational measurements.  In Fig. \ref{fig:csfrd1} we present the evolution of the TNG100-1 Cosmic star formation rate density alongside the observations of \citet{driver2018} and SDSS DR7 discussed in section \ref{sect:Obs}. We show that TNG100-1 is doing well against observations at $z < 1.4$. We note that the TNG model was tuned to do so, to surpass its successor original Illustris model that failed to reproduce the CSFRD at low redshifts. Besides the severe improvements, TNG100-1 implies higher values than observations at $z > 1.4$ and the TNG300-1 run performs better at earlier epochs with respect to the observations of \citet{driver2018}. We show that the agreement of TNG300-1 with high redshift observations is driven by resolution effects (Fig. \ref{fig:csfrd2}) and that the strong convergence test is not fulfilled for the TNG100-1 and TNG300-1 CSFRDs confirming that the problem of resolution effects go beyond the z $ \sim $ 0 star formation rate function. We note that \citet{Pillepich2018} performed resolution tests between simulations that adopted a 25 Mpc box and showed as well that higher resolutions resulted in higher cosmic star formation rate densities in the TNG model. We also note that similar problems would be found in most cosmological simulations including EAGLE and are not only specific for TNG. Three serious concerns arise for cosmological simulations besides their great improvements in the last 10 years from our analysis: 
\begin{itemize}
\item 1) Current state-of-the-art cosmological simulations can reproduce a range of observations (e.g. SFRF) mostly because there is a proper tuning of the parameters of their model at the {\it adopted resolution}. If the same model is run in lower resolution (regardless if it offers better statistics due to a larger box-size) it produces galaxies with different properties (e.g. lower star formation rates). This brings the question: Is the model successful at the reference simulation (e.g. TNG100-1) for physical reasons? Or is it successful only for the adopted resolution and due to the tuning of parameters? \\

\item 2)  Good statistics of rare high star-forming galaxies are possible to be achieved in large box simulations (e.g. 205 Mpc). However, the large volume cannot alone validate a simulation to be used as a predictive tool if we need to re-scale the properties and statistics of its galaxies due to limited resolution by the same level as the tension between observational studies. For example, the re-scaling needed between TNG300-1 and TNG100-1 SFRFs to bring them into an agreement at the high star-forming end is almost equal to the discrepancy between different star formation rate indicators \citep{Katsianis2017}, so TNG300-1 cannot be used as a predictive simulation at the high star-forming end to distinguish between different observational studies and be a guide for future surveys.  \\

\item 3) Future simulations that will achieve higher resolutions and adopt current state-of-the-art models (e.g. TNG or EAGLE) will without doubts need to re-scale their current parameters for sub-grid Physics and Feedback to reproduce some observables. However, with proper re-scaling any of the above models at the adopted reference resolution will be able to reproduce critical constraints like the stellar mass function and the evolution of the cosmic star formation rate density. Which are the observables that can determine the success of a model? Should any comparisons be mostly qualitative instead of quantitative? \\

\end{itemize}
In Fig. \ref{fig:image22}, we demonstrate that the peak of sSFRF changes among different resolutions, since the TNG100-1 run has higher normalization by 2 times with respect the TNG100-2 configuration (left panel of Fig. \ref{fig:image22}).

\section{The Intrinsic local Cosmic SFR density}
\label{appen.csfrd}
\begin{figure} 
   \centering
   \includegraphics[width=0.7\textwidth]{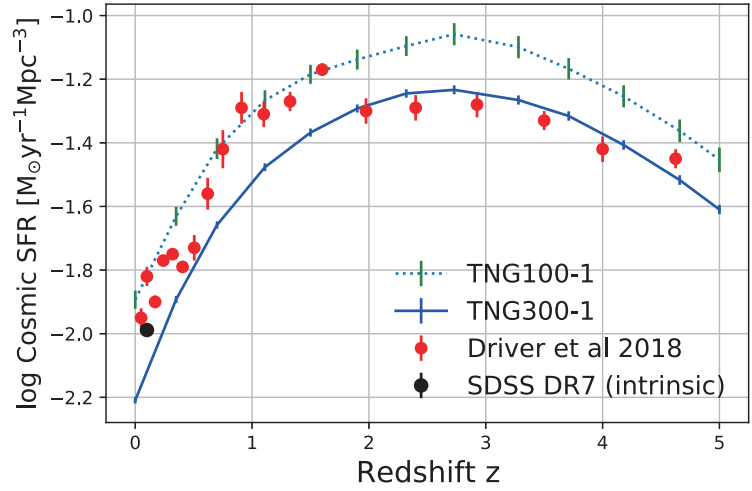}
   \caption{Cosmic star formation density: the green dotted line represents TNG100-1, blue solid line TNG300-1, magenta dots are the observations of \citet{driver2018} and the black dot is the intrinsic CSFRD of SDSS DR7, the error bars are obtained by Jacknife method.} 
   \label{fig:csfrd}
   \end{figure}

To obtain the cosmic SFR density of the local Universe, a  \citet{1976ApJ...203..297S} function is adopt to fit the inferred intrinsic SFRF shown in Fig.~\ref{Fig1}: 
\begin{equation}
  \rm \frac{d\phi}{dSFR} = \phi_{\star}\left(\frac{SFR}{SFR^\star}\right)^{\alpha}e^{-SFR/SFR^\star}\frac{1}{SFR^\star}
\label{eq:schechter}
\end{equation}
where $\alpha$ is the power-law slope of the low star-forming end and ${\rm SFR^\star}$ marks the characteristic SFR when the function shape transits from power-law to exponential cutoff. The $\phi_{\star}$ is the function amplitude at ${\rm SFR^\star}$. We only fit the inferred intrinsic SFRF where we consider is complete, i.e., SFRs $\ge 10^{-1.5}$ ${\rm M_{\odot} yr^{-1}}$. The best-fit parameters we obtained are $\phi_{\star} = 2.61\times10^{-3} \pm 9.49\times 10^{-5}$ ${\rm Mpc^{-3}}$, ${\rm SFR^\star = 2.89 \pm 0.07 }$ ${\rm M_{\odot}yr^{-1}}$, and $\alpha = -1.34 \pm 0.0115$. The intrinsic local Cosmic SFR density is therefore $9.74\times 10^{-3} \pm 4.51\times10^{-4}$ ${\rm M_\odot yr^{-1}Mpc^{-3}}$, as shown in Fig.~\ref{fig:csfrd} and Fig.~\ref{fig:csfrd22}. We report the above in order to facilitate parameter studies of the star formation rate function \citep{smit12,Tacchella2013} and cosmic star formation rate density \citep{Madau2014,Davies2016}.

\end{document}